\documentclass[aps,prl,oneside,twocolumn]{revtex4-1}
\usepackage{amsmath}
\usepackage{esint}
\usepackage{amssymb}
\usepackage{graphicx}
\usepackage{mathtools}

\begin{document}

\draft \title{Static hyperpolarizability of space-fractional quantum systems}
\author{Nathan J. Dawson}
\affiliation{Department of physics, College of The Bahamas, Nassau, Bahamas}
\email{dawsphys@hotmail.com}
\affiliation{Department of physics and astronomy, Washington State University, Pullman, WA, 99164, USA}
\date{\today}

\begin{abstract}
The nonlinear response is investigated for a space-fractional quantum mechanical system subject to a static electric field. Expressions for the polarizability and hyperpolarizability are derived from the fractional Schr\"{o}dinger equation in the particle-centric view for a three-level model constrained by the generalized Thomas-Reiche-Kuhn sum rule matrix elements. These expressions resemble those for a semi-relativistic system, where the reduction of the maximum linear and nonlinear static response is attributed to the functional dependence of the canonical position and momentum commutator. As examples, a clipped quantum harmonic oscillator potential and slant well potential are studied. The linear and first nonlinear response to the perturbing field are shown to decrease as the space fractionality is moved further below unity, which is caused by a suppression of the dipole transition moments. These results illustrate the importance of dimensionality and the order of the kinetic momentum operator which affect the strength of a system's optical response.
\end{abstract}

\maketitle

\section{Introduction}

The nonlinear electronic response to a perturbing electric field can be derived from fundamental quantum mechanics applied to the standard Schr\"{o}dinger equation.\cite{orr71.01} Sixteen years ago Kuzyk \cite{kuzyk00.01} discovered limits to the nonlinear optical coefficients based on the Hamiltonian found in the standard Schr\"{o}dinger equation with a real, spatial-dependent potential. Since that time, there has been an observed gap between the fundamental limit and the dipolar response coefficients from numerical optimizations as well as a larger gap between the fundamental limit and those coefficients determined experimentally from real molecules.\cite{kuzyk13.01} Many new strategies have been employed by researchers in recent years to bridge both of these gaps such as using Monte Carlo simulations to optimize mechanical potentials,\cite{kuzyk08.01,shafe10.01} studying quantum graphs as nonlinear optical materials that correlate the topology to the response,\cite{lytel13.01,lytel13.02} placing prongs on nanowires and conjugated molecules that cause a phase disruption,\cite{lytel15.01} using conjugation as a means to scale molecules and increase the linear and nonlinear coefficients,\cite{slepk04.01,perez07.01,cleuv14.01} and synthesizing twisted molecules.\cite{kang05.01,shi15.01}

Investigators have recently proposed the necessity for more exotic Hamiltonians due to the gap between the numerical simulations and the fundamental limit.\cite{watki12.01,shafe13.01,burke13.01} It was previously shown using the relativistically corrected Thomas-Reiche-Kuhn (TRK) sum rule \cite{cohen98.01,sinky06.01} that the fundamental limits to the nonlinear-optical coefficients were smaller than the nonrelativistic limits due to the inclusion of higher-order momentum operators reducing the total oscillator strength.\cite{dawson15.01} The space-fractional Schr\"{o}dinger equation \cite{laski02.01} has a kinetic energy term that acts on the spatial coordinates via the Riesz fractional derivative.\cite{riesz49.01} The canonical momentum acts as a fractional derivative operator on functions of position which changes the commutation relations and suppresses the net oscillator strength. The limit derived from the standard Schr\"{o}dinger equation is used to define the apparent intrinsic hyperpolarizability to compare the hyperpolarizablity from the fractional Schr\"{o}dinger equation with standard systems in linear space. Because the limit to the hyperpolarizability from the fractional Schr\"{o}dinger equation depends on the fractionality of space, it is shown to be more appropriate to study the apparent intrinsic hyperpolarizability.

Some applications of the fractional Schr\"{o}dinger equation to fundamental physics have previously been investigated such as calculations of the observed baryon spectrum at low energy,\cite{herrm07.01} an analytical model for ground-state band spectra of even-even nuclei,\cite{herrm10.01} liquid helium in nanoporous media,\cite{tayur12.01} and superfluids in capillary tubes as realizations of probability transport of relativistic particles.\cite{wei16.01} The results from this study could be used to validate and fine tune the parameters of some proposed applications of the fractional Schr\"{o}dinger equation via the sensitivity of nonlinear processes to dimensionality.

%------------------------------------------------

\section{Theory}

The nonlinear optical response of a system described by the fractional Schr\"{o}dinger equation has yet to be considered. Studying the sensitivity of the hyperpolarizability to the degree of spatial fractionality gives new insights to the observed response of quantum systems with space-fractional operators. The fractional Schr\"{o}dinger equation was first derived sixteen years ago by Laskin through a generalization of the path integral formulation via a L\'{e}vy-type stochastic motion.\cite{laski00.01,laski00.02} The space-fractional Schr\"{o}dinger equation for a single particle is given as
\begin{equation}
\hat{H}_\alpha \psi = E \psi ,
\label{eq:fracschrod}
\end{equation}
where the fractional Hamiltonian is
\begin{equation}
\hat{H}_{\alpha}\left(\boldsymbol{x},\boldsymbol{p}\right) = \frac{\left|\boldsymbol{p}\right|^{\beta}}{2m} + V\left(\boldsymbol{x}\right) .
\label{eq:hamiltfull}
\end{equation}
Here, $\psi$ is the wavefunction, $E$ is the energy, $m$ is the rest mass, and $V\left(\hat{x}\right)$ is the potential. The parameter $\beta = 2\alpha$ is real with $\alpha \rightarrow 1$ corresponding to the standard Schr\"{o}dinger equation.

It is often convenient to study one-dimensional systems that reduce a highly indexed tensor to a single diagonal component. The Hamiltonian along a single fractionalized direction is
\begin{equation}
\hat{H}_{\alpha}\left(\hat{x},\hat{p}\right) = \frac{\hat{p}^2}{2m} + V\left(\hat{x}\right) ,
\label{eq:hamilt}
\end{equation}
which is given in terms of the fractional coordinate $\hat{x}$. The canonical position and momentum are respectively given in terms of the fractional spatial coordinate and space-fractional derivative,
\begin{eqnarray}
\hat{x} &=& \left(\frac{\hbar}{mc}\right)^{1-\alpha} \left|x\right|^\alpha \mathrm{sign}\left(x\right), \label{eq:canonpos} \\
\hat{p} &=& -i mc \left(\frac{\hbar}{mc}\right)^{\alpha} \frac{\partial^\alpha}{\partial x^\alpha} , \label{eq:canonmom}
\end{eqnarray}
where $\partial^\alpha/\partial x^\alpha$ is the fractional derivative.\cite{herrm14.01} The constants in Eqs. \ref{eq:canonpos} and \ref{eq:canonmom} maintain proper dimensions. Note that we have used the well-known constants for the rest mass of an electron $m$, speed of light in vacuum $c$, and reduced Planck constant $\hbar$ to maintain the consistency of dimensions; however, any constants can be used that preserve the dimensions.

Note that there are several definitions of the fractional derivative which are all correct for the range of order over which they are defined. For fractional derivatives that are of positive order and less than one, the fractional derivative definitions are typically denoted as left- and right-sided derivatives. This notation for the direction is based on the integer value at the extreme of the derivative's range, where the first-order derivative is an odd operator. For finite difference calculations of differential equations, the first-order derivative is often taken with a direction, \textit{e}.\textit{g}. upwinding, where the same is true for a fractional derivative of order $(0,1]$. The Riesz definition of the fractional derivative, \textit{i}.\textit{e}. $\left(-\nabla^2 \right)^{\beta/2}$ for $1<\beta\leq 2$ with $\beta = 2\alpha$ is the fractional derivative based on the even-ordered Laplace operator and used in the fractional Schr\"{o}dinger equation.\cite{laski02.01} Similar to the previous analogy, finite difference approximations of second-order differential equations are often performed with a central difference approximation of the second-order derivative, which can be used for fast numerical schemes to solve the one-dimensional standard Schr\"{o}dinger equation. This central difference approximation ensures symmetry of the Laplace operator and preserves Hermiticity. Likewise, fractionalizing the second-order derivative using the Riesz definition also preserves the Hermiticity regardless of whether the Riesz fractional derivative is given in integral or differential form.

It is well-known that the nonlinear optical coefficients for a system subject to a static field can be determined via time-independent perturbation theory of the standard Schr\"{o}dinger equation.\cite{boyd08.01} Likewise, the nonlinear response to a static field of a space-fractional quantum system may also be determined via time-independent perturbation theory. The Hamiltonian of the perturbed system may be written as $\hat{H}_\alpha = \hat{H}_{\alpha}^{\left(0\right)} + \hat{V}_{\alpha}^{\mathrm{pert}}$, where $H_{\alpha}^{\left(0\right)}$ is the right-hand-side of Eq. \ref{eq:hamiltfull} and
\begin{equation}
\hat{V}_{\alpha}^{\mathrm{pert}} = e {\cal E} \hat{x} .
\label{eq:perthamilt}
\end{equation}
Here, $e$ is the charge of an electron and ${\cal E}$ is a constant electric field. Note that the perturbing field to be ``turned on'' is taken as a constant in fractional space.

The fractional Schr\"{o}dinger equation is the generalization of a second-order differential equation. Because bound states exist for particles in potential wells, the generalization to quantum systems modeled with the fractional Schr\"{o}dinger equation greatly depends on the choice of the well location with respect to the origin for non-integer $\alpha$. Thus, a particle-centric model is adopted for the remainder of this study, where the origin is located at the ground state expectation value.

Following the same scheme as Sakurai,\cite{sakur94.01} the second- and third-order energy shifts from time-independent perturbation theory are
\begin{eqnarray}
E^{\left(2\right)} = \left.\displaystyle\sum_{k}\right.^\prime  \frac{\left(\hat{V}_{\alpha}^{\mathrm{pert}} \right)_{0k} \left(\hat{V}_{\alpha}^{\mathrm{pert}} \right)_{k0}}{E_{k0}} ,
\label{eq:2ndcorr}
\end{eqnarray}
and
\begin{equation}
E^{\left(3\right)} = \displaystyle\left. \sum_{k,\ell}\right.^\prime  \frac{\left(\hat{V}_{\alpha}^{\mathrm{pert}} \right)_{0k} \left(\overline{V}_{\alpha}^{\mathrm{pert}} \right)_{k \ell} \left(\hat{V}_{\alpha}^{\mathrm{pert}} \right)_{\ell 0}}{E_{k0} E_{\ell0}}  ,
\label{eq:3rdcorr}
\end{equation}
where $E_{ij} = E_{i} - E_{j}$ and the prime denotes the sum over all states \textit{except} the ground state. Shorthand notation was introduced in Eqs. \ref{eq:2ndcorr} and \ref{eq:3rdcorr} where $\hat{{\cal O}}_{ij} = \left\langle i^{\left(0\right)} \right| \hat{{\cal O}} \left| j^{\left(0\right)} \right\rangle$ and $\overline{{\cal O}}_{ij} = \hat{{\cal O}}_{ij} - \delta_{ij} \hat{{\cal O}}_{00}$ with $\delta$ representing the Kronecker delta function.

The ground state energy is given as $E_0 = E_{0}^{\left(0\right)} + E_{0}^{\left(1\right)} + E_{0}^{\left(2\right)} + E_{0}^{\left(3\right)} + \cdots$ and the $n$th order scalar response to the static field is given by
\begin{equation}
\kappa^{\left(n\right)} = \displaystyle \frac{{\cal P}}{\left(n\right)!} \displaystyle \frac{\partial^n}{\partial {\cal E}^n} E_0\left({\cal E}\right) ,
\label{eq:kappan}
\end{equation}
where ${\cal P}$ is the permutation operator and the dipole moment is defined as $d = d^{\left(0\right)} + \sum_{n = 1} \kappa^{\left(n\right)}{\cal E}^n$. Note that the number of permutations of the input fields is equal to $\left(n+1\right)!$. The permanent dipole term, $d^{\left(0\right)}$, has been treated separately due to the particle-centric view placing the ground state expectation value of the particle at the origin.

Substituting Eq. \ref{eq:perthamilt} into Eqs. \ref{eq:2ndcorr} and \ref{eq:3rdcorr} shows that the $n$th-order correction to the ground state energy corresponds to an $n$th-order field term. The respective scalar polarizability, and scalar hyperpolarizability are given as
\begin{equation}
\kappa^{\left(1\right)} = 2 \displaystyle e^2 \displaystyle \left.\sum_{k} \right.^\prime \frac{\left(\hat{x}\right)_{0k} \left(\hat{x}\right)_{k0}}{E_{k0}} , \label{eq:polar}
\end{equation}
and
\begin{equation}
\kappa^{\left(2\right)} = 3 \displaystyle e^3 \displaystyle \left.\sum_{k,\ell} \right.^\prime \frac{\left(\hat{x}\right)_{0k} \left(\overline{x}\right)_{k \ell} \left(\hat{x}\right)_{\ell 0}}{E_{k0} E_{\ell 0}} .
\label{eq:hyperpol}
\end{equation}
The bar operator in Eq. \ref{eq:hyperpol} may be ignored in this treatment because the origin is placed at the ground state expectation value.

%An intrinsic value compare the nonlinear optical response of different systems for the same values of $\alpha$. To simplify the result, only the diagonal tensor component of the polarizability and hyperpolarizability for a one-dimensional system will be considered, \textit{i}.\textit{e}. $\kappa_{11}^{\left(1\right)}$ and $\kappa_{111}^{\left(2\right)}$, and henceforth, the coordinate subscripts will be removed.

The maximum value of the hyperpolarizability is well-known for $\alpha = 1$.\cite{kuzyk00.01,kuzyk13.01} The commutation relation between the canonical position and momentum operators, $\left[\hat{x},\hat{p}\right]$, does not necessarily reduce to the constant $i\hbar$, where properties from integer calculus such as the Leibniz rule and chain rule do not take the same form in fractional calculus. It follows that the TRK sum rule \cite{thoma25.01,reich25.01,kuhn25.01} for a space-fractional quantum system also has a more general form than that obtained from a mechanical Hamiltonian in the standard Schr\"{o}dinger equation. In general, we may evaluate the fractional TRK sum rule for a single particle by evaluating the elements of the commutator matrix for known wavefunctions,
\begin{equation}
\left\langle k \right| \left[\hat{x}, \left[\hat{H}_{\alpha},\hat{x}\right]\right] \left| \ell \right\rangle = \left\langle n \right| \left( 2\hat{x} \hat{H}_\alpha \hat{x} - \hat{H}_\alpha \hat{x}^2 - \hat{x}^2 \hat{H}_\alpha \right) \left| \ell \right\rangle .
\label{eq:TRKfracPre}
\end{equation}
The $\left(k,\ell\right)$ element on the right-hand-side of Eq. \ref{eq:TRKfracPre} may be rewritten using closure to give the fractional TRK sum rule as
\begin{align}
& \displaystyle \sum_{q = 0}^\infty \left(\hat{x}\right)_{kq} \left(\hat{x}\right)_{q\ell} \left[E_q - \frac{1}{2} \left(E_k + E_\ell\right)\right] \nonumber \\
&= \displaystyle \frac{\hbar^2}{4 m} \, \lambda\left(\alpha,k,\ell\right) ,
\label{eq:TRK}
\end{align}
where
\begin{equation}
\lambda\left(\alpha,k,\ell\right) = \left\langle k \right| \left( \frac{1}{2} \hat{\xi}^2 \frac{\partial^{2\alpha}}{\partial x^{2\alpha}} + \frac{1}{2} \frac{\partial^{2\alpha}}{\partial x^{2\alpha}}  \hat{\xi}^2 - \hat{\xi} \frac{\partial^{2\alpha}}{\partial x^{2\alpha}} \hat{\xi} \right) \left| \ell \right\rangle
\label{eq:lambdaparam}
\end{equation}
and $\hat{\xi} = \left|x\right|^\alpha \mathrm{sign}\left(x\right)$.

It is well-known that the maximum hyperpolarizability can be achieved from the TRK sum rule by allowing an ${\cal N}$-level system to approach an $\left({\cal N}-1\right)$-level system, which is achieved by forcing the highest transition energy to approach infinity. A previous study on multi-level systems predicts this maximum to increase as the $\sqrt{N}$, where these systems achieve the maximum by assuming degenerate low-lying states while letting the highest state's energy approach infinity.\cite{shafe13.01} There are currently no known potentials that allow such energy-level separations. The maximum hyperpolarizability derived from the TRK sum rule with only three levels has traditionally been regarded as the fundamental limit and is the common method used to compare hyperpolarizabilities. %This study adopts the intrinsic polarizability and hyperpolarizability based on the three-level ansatz for space-fractional quantum systems.

The largest fractional transition moment occurs when all of the oscillator strength is in the transition between the ground state to first excited state. This is observed from the $\left(0,0\right)$ TRK sum rule,
\begin{equation}
E_{10}\left|\left(\hat{x}\right)_{10}\right|^2 = \frac{\hbar^2}{2 m} \lambda\left(\alpha,0,0\right)  - \displaystyle \sum_{q = 2}^{\infty} E_{q0}\left|\left(\hat{x}\right)_{q0}\right|^2 .
\label{eq:00TRKPre}
\end{equation}
The largest, positive transition moment between the ground state and first excited state that is allowed by the TRK sum rule is given by
\begin{equation}
\hat{x}_{10}^{\mathrm{max}} = \frac{\hbar }{\sqrt{2 m E_{10}}} \sqrt{\lambda\left(\alpha,0,0\right)} .
\label{eq:xmax}
\end{equation}
Note that because all transition moments are assumed to be real, and because Hermiticity is preserved by the fractional Riesz operator, the fractional transition moments must have the property $\left(\hat{x}\right)_{ij} = \left(\hat{x}\right)_{ji}$.

\begin{figure*}[t]
\centering\includegraphics[scale=1]{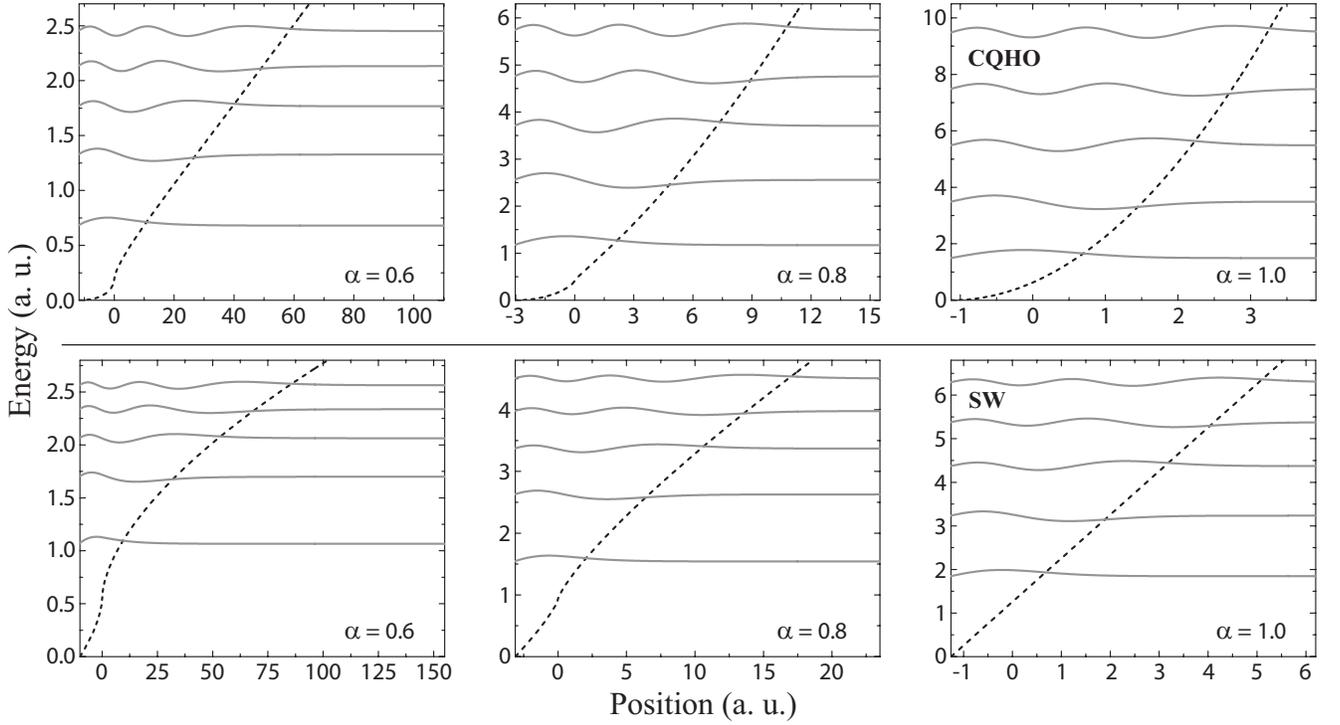}
\caption{The potential well (dashed) and first five wavefunctions (solid) for a CQHO (top) and SW (bottom) potential for different $\alpha$.}
\label{fig:FracPot}
\end{figure*}

To reduce the number of parameters, it is useful to define the transition energy ratio, $E = E_{10}/E_{20}$, and the ratio of the ground state to first excited state relative to the maximum value, $\hat{X} = \left|\hat{x}_{10}\right| / \hat{x}_{10}^{\mathrm{max}}$. Multiplying both sides of the latter equation by $\hat{x}_{10}^{\mathrm{max}}$ gives
\begin{equation}
\hat{x}_{10} = \frac{\hbar}{\sqrt{2 m E_{10}}} \hat{X} \sqrt{\lambda\left(\alpha,0,0\right)} .
\label{eq:x01}
\end{equation}
The remaining transition moments may be expressed in terms of $X$, $E$, and $\lambda$ from the $\left(0,1\right)$, $\left(0,2\right)$, $\left(0,0\right)$, and $\left(1,1\right)$ elements of the TRK sum rule. The expressions for these sum-rule-constrained transition moments for a three-level model are
\begin{align}
\hat{x}_{20} &= \frac{\hbar}{\sqrt{2 m E_{10}}} \sqrt{E\left(1-\hat{X}^2\right)} \sqrt{\lambda\left(\alpha,0,0\right)} \label{eq:x02} \\
\hat{x}_{12} &= \frac{\hbar}{\sqrt{2 m E_{10}}} \sqrt{\frac{E}{1-E}}\sqrt{\hat{X}^2 \lambda\left(\alpha,0,0\right) + \lambda\left(\alpha,1,1\right)} \label{eq:x12} \\
\overline{x}_{11} &= \frac{\hbar}{\sqrt{2 m E_{10}}}\Bigg[ \frac{E-2}{\sqrt{1-E}} \frac{\sqrt{1-\hat{X}^2}}{\hat{X}} \sqrt{\hat{X}^2 \lambda\left(\alpha,0,0\right) + \lambda\left(\alpha,1,1\right)} \nonumber \\
&- \frac{1}{\hat{X}}\frac{\lambda\left(\alpha,1,0\right)}{\sqrt{\lambda\left(\alpha,0,0\right)}}\Bigg] \label{eq:x11} \\
\overline{x}_{22} &= \frac{\hbar}{\sqrt{2 m E_{10}}}\Bigg[ \frac{1-2E}{\sqrt{1-E}} \frac{\hat{X}}{\sqrt{1-\hat{X}^2}} \sqrt{\hat{X}^2 \lambda\left(\alpha,0,0\right) + \lambda\left(\alpha,1,1\right)} \nonumber \\
&- \sqrt{\frac{E}{1-\hat{X}^2}} \frac{\lambda\left(\alpha,2,0\right)}{\sqrt{\lambda\left(\alpha,0,0\right)}}\Bigg] . \label{eq:x22}
\end{align}

Substituting these transition moment expressions into Eqs. \ref{eq:polar} and \ref{eq:hyperpol}, the space-fractional polarizability and hyperpolarizability for the three-level model reduce to
\begin{equation}
\kappa^{\left(1\right)} = \frac{e^2 \hbar^2}{m E_{10}^{2}} \left[\hat{X}^2 + E^2 \left(1-\hat{X}^2\right)\right] \lambda\left(\alpha,0,0\right)
\label{eq:3Lpol}
\end{equation}
and
\begin{align}
\kappa^{\left(2\right)} &= \frac{3}{2} \frac{e^3 \hbar^3}{\sqrt{2 m^3 E_{10}^{7}}} \Bigg[ \hat{X} \sqrt{1-\hat{X}^2} \left(1-E\right)^{3/2} \left(2 + 3E +2 E^2 \right) \nonumber \\
&\times \lambda\left(\alpha,0,0\right) \sqrt{\hat{X}^2 \lambda\left(\alpha,0,0\right) + \lambda\left(\alpha,1,1\right)} \nonumber \\
&- \hat{X} \sqrt{\lambda\left(\alpha,0,0\right)} \lambda\left(\alpha,1,0\right) \nonumber \\
&- \sqrt{1 - \hat{X}^2} E^{7/2} \sqrt{\lambda\left(\alpha,0,0\right)} \lambda\left(\alpha,2,0\right) \Bigg] .
\label{eq:3Lhyper}
\end{align}
When $\alpha\rightarrow 1$, the Hamiltonian from the standard Schr\"{o}dinger equation is retrieved, where $\lambda\left(\alpha\rightarrow1,i,i\right)\rightarrow 1$ and $\lambda\left(\alpha\rightarrow1,i,j\right)\rightarrow 0$ for $i\neq j$. The maximum polarizability for the case of $\alpha\rightarrow 1$ occurs when $\hat{X} = 1$ for any value of $E$. The maximum hyperpolarizability when $\alpha\rightarrow 1$ occurs when $\hat{X} = \sqrt[-4]{3}$ and $E\rightarrow 0$ which gives
\begin{equation}
\kappa_{\mathrm{max},\alpha\rightarrow 1}^{\left(2\right)} = \sqrt[4]{3} e^3 \hbar^3 \sqrt{\frac{N^3}{m^3 E_{10}^{7}}} .
\label{eq:3Lhypermax}
\end{equation}
For the case of non-integer $\alpha$, the TRK sum rule elements depend on the wavefunctions. Thus, the generalized maximum of the hyperpolarizability, $\kappa_{\mathrm{max},\alpha}^{\left(2\right)}$, also depends on $\alpha$ as well as the wavefunctions.

The diagonal matrix elements $\lambda\left(\alpha,0,0\right)$ and $\lambda\left(\alpha,1,1\right)$ can change the magnitude of the response, where these values approach unity when $\alpha$ approaches unity. This vanishing of the wavefunction dependence of the fundamental limits of nonlinear optical coefficients for $\alpha = 1$ is due to the TRK sum rule resulting in a constant for the standard Schr\"{o}dinger equation with a mechanical Hamiltonian. The momentum operator for $\alpha \neq 1$ in the kinetic energy that results in modified limits to the optical response illustrate the importance of the momentum operator in space-fractional systems. Thus, it appears that the limit for $\alpha = 1$ is a special case that allows for such a simple expression with no wavefunction dependence, where systems that can be described by the space-fractional Schr\"{o}dinger equation can have nonlinear optical coefficients that are much smaller.

The terms that contain the parameters $\lambda\left(\alpha,1,0\right)$ and $\lambda\left(\alpha,2,0\right)$ in Eq. \ref{eq:3Lhyper} directly subtract from the non-vanishing term in the $\alpha\rightarrow 1$ limit. Large enough values of the off-diagonal $\lambda$ matrix can give a negative result, although it is noted that these off-diagonal elements can be of either sign and are expected to be negligible. Because the hyperpolarizability is an odd coefficient, the value can either be negative or positive depending upon the chosen positive direction of the axis relative to the shape of the potential. For large, positive, off-diagonal $\lambda$ parameters, it appears that this sign could flip such that large positive coefficients can become large negative coefficients, which appears to be an unreasonable shift.

A similar strange behavior of the limit to the hyperpolarizability under the three-level ansatz has also been observed for systems with relativistic kinetic energies. Here, the hyperpolarizability was also derived from the TRK sum rule under the three-level ansatz, where many additional parameters are introduced from higher-order momentum operators in the kinetic energy via the Foldy-Wouthuysen transformation. A more in-depth study of the functional forms for larger numbers of included states (and the corresponding $\lambda$ parameters) may be necessary in the future to fully understand the consequences of forcing a three-level model to arrive at Eq. \ref{eq:3Lhyper} for space-fractional quantum systems.

Note that this formulation only considers the fractional Schr\"{o}dinger equation using the Riesz fractional derivative of order $(1,2]$. When $\alpha$ is larger than unity ($\beta > 2$), regardless of being an integer value, the canonical momentum depends on the kinetic momentum to a power greater than unity; however a definition of the fractional differential operator different from the Riesz definition is necessary to describe the phenomena. Several new definitions of higher-order fractional derivatives have recently been introduced based on finite differences.\cite{herrm09.01} These definitions are based on the highest integer derivative associated with the upper bound of $\alpha$. For example, for a fractional derivative $\partial^\beta/\partial x^\beta$ with a range of $0<\beta\leq 3$, the definition is based on the finite difference form of $\partial^3/\partial x^3$, which is not symmetric. Likewise, the definition of the fractional derivative with range $0<\beta\leq 4$ is based on the central difference of $\partial^4/\partial x^4$, which is an even operator resulting in a Hermitian matrix for real $x$. Thus, we would expect that space-fractional systems generalizing the one-dimensional standard Schr\"{o}dinger equation to have momentum operators than never exceed $\alpha = 1$. Therefore, it follows that the fractional Laplace operator resulting from the $\left|\boldsymbol{p}\right|^{2\alpha} / 2m$ kinetic energy term must be described by a fractional differential operator based on the integer operator of second order, which places an upper bound on the range of the fractional parameter. If a system is described by a higher-order fractional derivative, then it most likely stems from a kinetic energy term that is not a generalization of the classical kinetic energy.

\section{Results and discussion}

The hyperpolarizability of asymmetric potentials is non-zero, and unlike the polarizability, its strength is sensitive to intermediate transitions. To satisfy the asymmetry of the system, two half potentials were chosen with a single bound particle. The fractional clipped quantum harmonic oscillator (CQHO) is defined as
\begin{equation}
V\left(\hat{x}\right) = \begin{dcases} \frac{1}{2} m \omega^2 \left( \hat{x} - \hat{b} \right)^2 & \mathrm{for} \quad \hat{x} > \hat{b} \\
\infty & \mathrm{for} \quad \hat{x} \leq \hat{b} , \end{dcases}
\label{eq:CQHO}
\end{equation}
where $\omega$ is the angular frequency. The fractional offset of the potential in Eq. \ref{eq:CQHO} is
\begin{equation}
\hat{b} = \left(\frac{\hbar}{mc}\right)^{1-\alpha} \left|b\right|^\alpha \mathrm{sign}\left(b\right) ,
\label{eq:bhat}
\end{equation}
where $b$ is the offset to the standard CQHO potential and ensures the particle-centric view such that $\hat{x}_{00} = 0$. The other asymmetric potential chosen for this study is the fractional slant well (SW) given by
\begin{equation}
V\left(\hat{x}\right) = \begin{dcases} A \left( \hat{x} - \hat{b} \right) & \mathrm{for} \quad \hat{x} > \hat{b} \\
\infty & \mathrm{for} \quad \hat{x} \leq \hat{b} . \end{dcases}
\label{eq:SW}
\end{equation}

The potential and first five wavefunctions of the fractional CQHO and SW potentials are shown in Fig. \ref{fig:FracPot} for increasing $\alpha$. Because a finite difference scheme is used to approximate the Hamiltonian, the fractional quantum Riesz derivative is not represented in integral form. Rather the finite difference representation is obtained by fast approximation using the half sum of the left- and right-sided Caputo derivatives.\cite{podlu09.01} The CQHO potential as a function of the canonical position approaches the SW as a function of the position $x$; however, the canonical momentum decreases the order of the differential operator. Note that the potential increasingly scales in $x$ as $\alpha$ decreases due to the speed of light being taken as the inverse of the fine structure constant. Arbitrary constants can also be used to preserve the dimensions of the fractional Schr\"{o}dinger equation that adjust this scaling without affecting the intrinsic response.

The intrinsic hyperpolarizability is used to compare the nonlinear response of a quantum system to other systems of arbitrary scale, potential shape, and class that are all described by the standard Schr\"{o}dinger equation. The maximum hyperpolarizability for a fractional quantum systems has additional dependencies given by the $\lambda$ matrix which makes the true intrinsic hyperpolarizability, $\kappa_{\mathrm{int}}^{\left(2\right)} = \kappa^{\left(2\right)} / \kappa_{\mathrm{max},\alpha}^{\left(2\right)}$, less intuitive. A simple parameter that compares the nonlinear response of systems described by the fractional Schr\"{o}dinger equation to systems that are described by the standard Schr\"{o}dinger equation is defined as the apparent intrinsic hyperpolarizability, $\kappa_{\mathrm{app}}^{\left(2\right)} = \kappa^{\left(2\right)} / \kappa_{\mathrm{max},\alpha\rightarrow 1}^{\left(2\right)}$.

\begin{figure}[t]
\centering\includegraphics[scale=1]{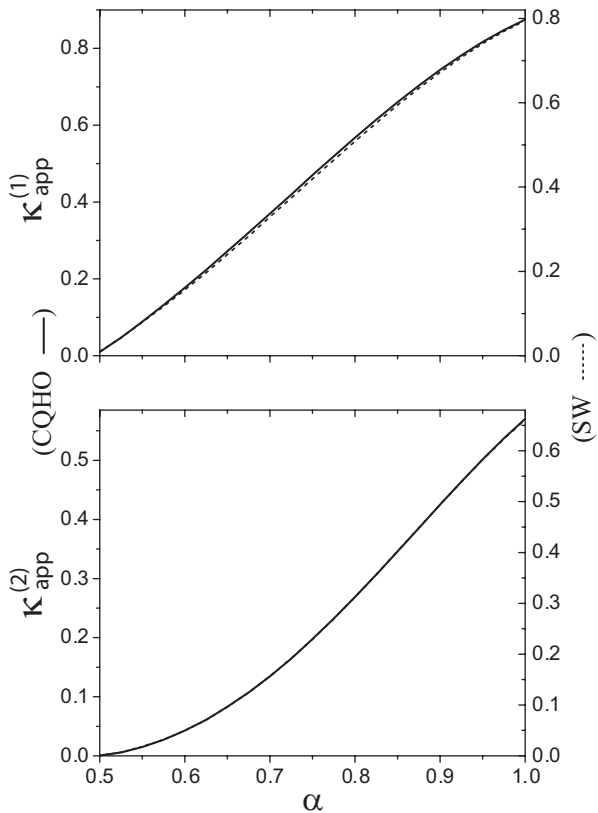}
\caption{The apparent intrinsic polarizability $\kappa_{\mathrm{app}}^{\left(1\right)}$ and apparent intrinsic hyperpolarizability $\kappa_{\mathrm{app}}^{\left(2\right)}$ for the CQHO ($\omega = 1$ a.u.) and  SW ($A = 1$ a.u.) potentials as a function of the fractional parameter, $\alpha$.}
\label{fig:fracvals}
\end{figure}

The apparent intrinsic polarizability and apparent intrinsic hyperpolarizability as a function of $\alpha$ are shown in Fig. \ref{fig:fracvals} for the CQHO and SW potentials. These values are calculated using the sum-over-states expressions. The origin can be determined below a specified threshold by solving the space-fractional Schr\"{o}dinger equation and updating the offset via an iterative process. The hyperpolarizability is more sensitive to the value of $\alpha$ than the polarizability, where even a small deviation from unity causes a significant decrease in the nonlinear response. Note that this sensitivity to $\alpha$ is minimized in the particle-centric view, where moving the origin in either direction away from $\hat{x}_{00}$ will cause the polarizability and hyperpolarizability to decrease at an even faster rate as $\alpha$ moves away from unity. Note that the symmetry about the origin is conserved for the fractional wave equation. Thus, the hyperpolarizability remains zero for a particle in a symmetric potential centered at $\hat{x}_{00} = 0$.

\begin{figure}[t]
\centering\includegraphics[scale=1]{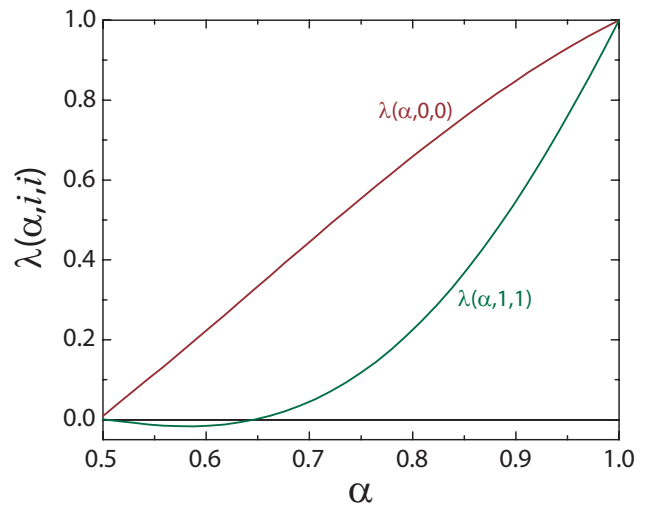}
\caption{The $\lambda\left(\alpha,0,0\right)$ and $\lambda\left(\alpha,1,1\right)$ elements as functions of $\alpha$ for the CQHO potential.}
\label{fig:lambda}
\end{figure}

The linear and nonlinear optical response trending to zero is of no surprise after the examination of the three-level model given by Eq. \ref{eq:3Lhyper}. The fractional dependent element $\lambda\left(\alpha,0,0\right)$ suppresses the linear response while both $\lambda\left(\alpha,0,0\right)$ and $\lambda\left(\alpha,1,1\right)$ rapidly dampen the nonlinear response. These diagonal parameters are shown in Fig. \ref{fig:lambda} for the CQHO potential, where the linear polarizability has a similar downward trend to $\lambda\left(\alpha,0,0\right)$. The hyperpolarizability is more sensitive to $\alpha$ because the TRK sum rule elements used to derive Eq. \ref{eq:3Lhyper} have additional $\lambda$ elements, where the expression derived from a mechanical Hamiltonian in the standard Schr\"{o}dinger equation is suppressed by the $\lambda\left(\alpha,0,0\right)$ element and function of both $\lambda\left(\alpha,0,0\right)$ and $\lambda\left(\alpha,1,1\right)$. The off-diagonal elements were negligibly small for all values of $\alpha$ and sensitive to small variations of the ground-state expectation value with respect to the origin.

\section{Conclusion}

The nonlinear-optical coefficients in response to a static external field were derived for systems described by the space-fractional Schr\"{o}dinger equation. The TRK sum rule elements were shown to depend on the fractional parameter $\alpha$ and the wavefunctions. The polarizability and hyperpolarizability expressions derived from the fractional TRK sum rule for a three-level model resulted in an expression with four wavefunction-dependent parameters. These expressions resemble those obtained from relativistic corrections, where a correlation between the fractional Schr\"{o}dinger equation and the relativistic Schr\"{o}dinger equation has recently been reported.\cite{wei15.01}

A particle-centric model was used for the fractional Schr\"{o}dinger equation with an asymmetric potential. The apparent intrinsic hyperpolarizability has been shown to decrease at a much faster rate than the apparent intrinsic polarizability as $\alpha$ decreases from unity. The sensitivity of the nonlinear response to the differential operator's fractional power may help explain why the hyperpolarizability is significantly lower in experiment than some intrinsic values estimated from scaling arguments for mesoscopic systems. Characterizing the linear and nonlinear processes of proposed fractional quantum systems provides a method to test the applicability of the fractional Schr\"{o}dinger equation to some systems. Some systems that strongly interact with their environment could be described by the space-fractional Schr\"{o}dinger equation with low linear and nonlinear optical response.
The approximate relationship between the space-fractional Schr\"{o}dinger equation and the relativistic Schr\"{o}dinger equation also predicts a lower limit for the nonlinear optical coefficients of bound particles with semi-relativistic kinetic energies. Indeed, the interaction between light and matter appears to only be strong in the $p^2$ kinetic momentum limit. This damping of the linear and nonlinear optical response could also be applied to the `dark matter' problem,\cite{olive03.01,berto05.01,berto16.01} where neutral matter composed of strongly interacting charged particles, baryonic or otherwise, may interact negligibly with the electromagnetic spectrum when bound particles move far away from the $p^2$ limit.

%\bibliographystyle{osajnl}
%\bibliography{\bibs}

\begin{thebibliography}{10}
\newcommand{\enquote}[1]{``#1''}

\bibitem{orr71.01}
B.~J. Orr and J.~F. Ward, \enquote{{Perturbation Theory of the Non-Linear
  Optical Polarization of an Isolated System},} Molecular Physics \textbf{20},
  513--526 (1971).

\bibitem{kuzyk00.01}
M.~G. Kuzyk, \enquote{{Physical Limits on Electronic Nonlinear Molecular
  Susceptibilities},} Phys. Rev. Lett. \textbf{85}, 1218--1221 (2000).

\bibitem{kuzyk13.01}
M.~G. Kuzyk, J.~P\'{e}rez-Moreno, and S.~Shafei, \enquote{{Sum rules and
  scaling in nonlinear optics},} Phys. Rep. \textbf{529}, 297--398 (2013).

\bibitem{kuzyk08.01}
M.~C. Kuzyk and M.~G. Kuzyk, \enquote{{Monte Carlo Studies of the Fundamental
  Limits of the Intrinsic Hyperpolarizability},} J. Opt. Soc. Am. B.
  \textbf{25}, 103--110 (2008).

\bibitem{shafe10.01}
S.~Shafei, M.~C. Kuzyk, and M.~G. Kuzky, \enquote{{Monte Carlo studies of the
  intrinsic second hyperpolarizability},} J. Opt. Soc. Am. B \textbf{27},
  1849--1856 (2010).

\bibitem{lytel13.01}
R.~Lytel and M.~G. Kuzyk, \enquote{{Dressed Quantum Graphs with Optical
  Nonlinearities Approaching the Fundamental Limit},} J. Nonlinear Opt. Phys.
  Mat. \textbf{22}, 1350041 (2013).

\bibitem{lytel13.02}
R.~Lytel, S.~Shafei, J.~H. Smith, and M.~G. Kuzyk, \enquote{{Influence of
  geometry and topology of quantum graphs on their nonlinear optical
  properties},} Phys. Rev. A \textbf{87}, 043824 (2013).

\bibitem{lytel15.01}
R.~Lytel, S.~M. Mossman, and M.~G. Kuzyk, \enquote{{Phase disruption as a new
  design paradigm for optimizing the nonlinear-optical response},} Opt. Lett.
  \textbf{40}, 4735--4738 (2015).

\bibitem{slepk04.01}
A.~D. Slepkov, F.~A. Hegmann, S.~Eisler, E.~Elliot, and R.~R. Tykwinski,
  \enquote{{The surprising nonlinear optical properties of conjugated polyyne
  oligomers},} J. Chem. Phys. \textbf{120}, 6807--6810 (2004).

\bibitem{perez07.01}
J.~P\'{e}rez-Moreno, Y.~Zhao, K.~Clays, and M.~G. Kuzyk, \enquote{{Modulated
  conjugation as a means for attaining a record high intrinsic
  hyperpolarizability},} Opt. Lett. \textbf{32}, 59--61 (2007).

\bibitem{cleuv14.01}
S.~V. Cleuvenbergen, I.~Asselberghs, W.~Vanormelingen, T.~Verbiest, E.~Franz,
  K.~Clays, M.~G. Kuzyk, and G.~Koeckelberghs, \enquote{{Optimizing the second
  hyperpolarizability with minimally parameterized potentials},} J. Mater.
  Chem. C \textbf{2}, 4533--4538 (2014).

\bibitem{kang05.01}
H.~Kang, A.~Facchetti, P.~Zhu, H.~Jiang, Y.~Yang, E.~Cariati, S.~Righetto,
  R.~Ugo, C.~Zuccaccia, A.~Macchioni, C.~L. Stern, Z.~Liu, S.-T. Ho, and T.~J.
  Marks, \enquote{Exceptional molecular hyperpolarizabilities in twisted
  p-electron system chromophores,} Angew. Chem. Int. Ed. \textbf{44},
  7922--7925 (2005).

\bibitem{shi15.01}
Y.~Shi, D.~Frattarelli, N.~Watanabe, A.~Facchetti, E.~Cariati, S.~Righetto,
  E.~Tordin, C.~Zuccaccia, A.~Macchioni, S.~L. Wegener, C.~L. Stern, M.~A.
  Ratner, and T.~J. Marks, \enquote{Ultra-high-response, multiply twisted
  electro-optic chromophores: Influence of p-system elongation and interplanar
  torsion on hyperpolarizability,} J. Am. Chem. Soc. \textbf{137}, 12521--12538
  (2015).

\bibitem{watki12.01}
D.~S. Watkins and M.~G. Kuzyk, \enquote{{Universal properties of the optimized
  off-resonant intrinsic second hyperpolarizability},} J. Opt. Soc. Am. B
  \textbf{29}, 1661--1671 (2012).

\bibitem{shafe13.01}
S.~Shafei and M.~G. Kuzyk, \enquote{{Paradox of the many-state catastrophe of
  fundamental limits and the three-state conjecture},} Phys. Rev. A
  \textbf{88}, 023863 (2013).

\bibitem{burke13.01}
C.~J. Burke, T.~J. Atherton, J.~Lesnefsky, and R.~G. Petschek,
  \enquote{{Optimizing the second hyperpolarizability with minimally
  parameterized potentials},} J. Opt. Soc. Am. B \textbf{30}, 1438--1445
  (2013).

\bibitem{cohen98.01}
S.~M. Cohen and P.~T. Leung, \enquote{{General formulation of the
  semirelativistic approach to atomic sum rules},} Phys. Rev. A \textbf{57},
  4994--4997 (1998).

\bibitem{sinky06.01}
H.~Sinky and P.~T. Leung, \enquote{{Relativistic corrections to a generalized
  sum rule},} Phys. Rev. A \textbf{74}, 034703 (2006).

\bibitem{dawson15.01}
N.~J. Dawson, \enquote{{Lowest-order relativistic corrections to the
  fundamental limits of nonlinear-optical coefficients},} Phys. Rev. A
  \textbf{91}, 013832 (2015).

\bibitem{laski02.01}
N.~Laskin, \enquote{{Fractional Schr\"{o}dinger equation},} Phys. Rev. E
  \textbf{66}, 056108 (2002).

\bibitem{riesz49.01}
M.~Riesz, \enquote{{L'int\'{e}grale de Riemann-Liouville et le probl\`{e}me de
  Cauchy},} Acta Math. \textbf{81}, 1--222 (1949).

\bibitem{herrm07.01}
R.~Herrmann, \enquote{{The fractional symmetric rigid rotator},} J. Phys. G
  \textbf{34}, 607--625 (2007).

\bibitem{herrm10.01}
R.~Herrmann, \enquote{{Common aspects of $q$-deformed Lie algebras and
  fractional calculus},} Physica A \textbf{389}, 4613--4622 (2010).

\bibitem{tayur12.01}
D.~A. Tayurskii and Y.~V. Lysogorskiy, \enquote{{Superfluid hydrodynamic in
  fractal dimension space},} J. Phys. Conf. Ser. \textbf{394}, 012004 (2012).

\bibitem{wei16.01}
Y.~Wei, \enquote{{Comment on ``Fractional quantum mechanics'' and ``Fractional
  Schr\"{o}dinger equation''},} Phys. Rev. E \textbf{93}, 066103 (2016).

\bibitem{laski00.01}
N.~Laskin, \enquote{{Fractional quantum mechanics and L\'{e}vy path
  integrals},} Phys. Lett. A \textbf{268}, 298--305 (2000).

\bibitem{laski00.02}
N.~Laskin, \enquote{{Fractional quantum mechanics},} Phys. Rev. E \textbf{62},
  3135--3145 (2000).

\bibitem{herrm14.01}
R.~Herrmann, \emph{{Fractional Calculus: an introduction for physicists}}
  (World Scientific Publishing Co. Pte. Ltd., New Jersey, 2014), 2nd ed.

\bibitem{boyd08.01}
R.~W. Boyd, \emph{{Nonlinear Optics}} (Academic Press, San Diego, 2008), 3rd
  ed.

\bibitem{sakur94.01}
J.~J. Sakurai, \emph{{Modern Quantum Mechanics}} (Addison-Wesley, Reading,
  1994), 3rd ed.

\bibitem{thoma25.01}
W.~Thomas, \enquote{{\"{U}ber die Zahl der Dispersionselektronen, die einem
  station\"{a}ren Zustand zugeordnet sind.(Vorlaufige Mitteilung)},}
  Naturwissenschaften \textbf{13}, 627--627 (1925).

\bibitem{reich25.01}
F.~Reiche and W.~Thomas, \enquote{{\"{U}ber die Zahl der Dispersionselektronen,
  die einem station\"{a}ren Zustand zugeordnet sind},} Z. Phys. \textbf{34},
  510--525 (1925).

\bibitem{kuhn25.01}
W.~Kuhn, \enquote{{\"{U}ber die Gesamtst\"{a}rke der von einem Zustande
  ausgehenden Absorptionslinien},} Z. Phys. \textbf{33}, 408--412 (1925).

\bibitem{herrm09.01}
R.~Herrmann, \enquote{{Higher order fractional derivatives},} arXiv:0906.2185
  (2009).

\bibitem{podlu09.01}
I.~Podlubny, A.~Chechkin, T.~Skovranek, Y.~Chen, and B.~M.~V. Jara,
  \enquote{{Matrix approach to discrete fractional calculus II: Partial
  fractional differential equations},} J. Comput. Phys. \textbf{228},
  3137--3153 (2009).

\bibitem{wei15.01}
Y.~Wei, \enquote{{Some solutions to the fractional and relativistic
  Schr\"{o}dinger equations},} Int. J. Theor. and Math. Phys. \textbf{5},
  87--111 (2015).

\bibitem{olive03.01}
K.~A. Olive, \enquote{{TASI Lectures on Dark Matter},} arXiv:astro-ph/0301505
  (2003).

\bibitem{berto05.01}
D.~Bertone, G.~Hooper and J.~Silk, \enquote{{Particle dark matter: evidence,
  candidates and constraints},} Phys. Rep. \textbf{405}, 279--390 (2005).

\bibitem{berto16.01}
G.~Bertone and D.~Hooper, \enquote{{A History of Dark Matter},}
  arXiv:1605.04909v2  (2016).

\end{thebibliography}

\end{document}